\documentclass[11pt]{article}
\usepackage{epsfig}
\usepackage[usenames]{color}
\usepackage{graphicx}
\usepackage{epstopdf}
\newtheorem{proposition}{Proposition}[section]
\newcommand{\bpr}{\begin{proposition}}
\newcommand{\epr}{\end{proposition}}

\newcommand{\beq}{\begin{equation}}
\newcommand{\eeq}{\end{equation}}
\newcommand{\bea}{\begin{eqnarray}}
\newcommand{\eea}{\end{eqnarray}}
\input epsf
\begin{document}  
\hfill \vbox{\hbox{}} 
\begin{center}{\Large\bf  Critical couplings and string tensions via lattice matching of 
RG decimations}\\[2cm] 
{X. Cheng\footnote{\sf e-mail: darktree@physics.ucla.edu} and E. T. Tomboulis\footnote{\sf e-mail: tomboulis@physics.ucla.edu}
}\\
{\em Department of Physics, UCLA, Los Angeles, 
CA 90095-1547}
%{\sf e-mail: tombouli@physics.ucla.edu}
\end{center}
\vspace{1cm}

\begin{center}{\Large\bf Abstract}
\end{center} 
We calculate critical couplings and string tensions in $SU(2)$ and $SU(3)$ pure lattice gauge theory 
by a simple and inexpensive technique of two-lattice matching of 
RG block transformations. The transformations are potential moving decimations generating 
plaquette actions with large number of group characters and exhibit rapid approach to a unique renormalized trajectory. 
Fixing the critical coupling $\beta_c(N_\tau)$ at one value of temporal lattice length $N_\tau$ 
by MC simulation, the critical couplings for any other value of $N_\tau$ are then obtained by lattice matching of the block decimations. We obtain $\beta_c(N_\tau)$ values over the range $N_\tau = 3 - 32$ and find agreement with MC simulation results to within a few percent in all cases. A similar 
procedure allows the calculation of string tensions   
with similarly good agreement with MC data.

\vfill 

\pagebreak 

\section{Introduction} 

In this paper we apply the renormalization group (RG) based technique of  ``lattice 
matching" via block transformations in pure lattice gauge theories. Instead of  implementing 
block transformations by means of renormalization group Monte Carlo (MCRG) 
methods, however, we employ explicit RG recursion relations of the ``potential moving" type. 
The block transformations (decimations) implemented by these recursions are of course approximate
but can, in principle, be systematically improved.   They turn out to be surprisingly effective for various purposes. 
Here we use lattice matching of these decimations to obtain critical couplings and string tensions 
for the gauge groups $SU(2)$ and $SU(3)$. 

Specifically, we start with one critical coupling obtained by MC simulation at a certain temporal 
lattice size $N_\tau$.  
%- this essentially sets scales. 
We then employ lattice matching of our decimations to compute the critical couplings $\beta_c(N_\tau)$ at other values of $N_\tau$. We find that 
these values agree with the values obtained by MC to within at most a few percent. 
This procedure then affords a rather inexpensive method for obtaining critical couplings over  
a wide range of lattice sizes. In fact, since the recursions are locally specified, there appear to be no intrinsic lattice size limitations in the method.

A different method for inexpensive estimation of critical couplings has recently been presented in \cite{Phetal}. In this approach the strong coupling expansion of the $SU(N)$ theory is used to derive a $3$-dimensional $Z(N)$ effective Polyakov loop action. The easily ascertained critical values of the effective action parameters can then be mapped back to the critical couplings of the original $SU(N)$ theory. It would be interesting to consider combining a RG decimation-based method such as the one presented here with that of \cite{Phetal} to extend the range where the latter can be applied. 

Our lattice matching of decimations can also be applied, by a closely analogous procedure, to the computation of string tensions for various values of $\beta$.  Again, the computed values are in very good agreement with the MC data. 

The paper is arranged as follows. Our RG transformations and the resulting decimation recursions are formulated in section  \ref{RGdec}. 
The lattice matching method and its application via our decimations to the computation of critical couplings and string tensions are outlined in section  \ref{LM}. Our numerical results are 
presented in section \ref{NumRes}. Some concluding remarks are given in section \ref{Concls}.

\section{RG decimations \label{RGdec}} 

An RG block transformation with scale factor $b$ maps a system on a lattice of spacing 
$a$ to a system on a lattice of spacing $ba$. The flow in the space of interactions (couplings) under successive block transformations defines the RG flow of the system. 
%(The structure (number of relevant directions, critical exponents) of fixed points of this flow is 
%universal but their location is not. The latter, as also the form of the effective action after each 
%transformation, are dependent on the precise definition of the transformation, i.e. the choice of %block variables. Criteria for a good choice include preservation of (quasi)locality of the resulting 
%action after each block step. )

Given the definition of some exact block transformation, its practical implementation, whether 
by analytical or numerical methods, generally involves some approximation or truncation. 
This is certainly the case if one aims at obtaining the effective action after each step. The 
standard method for implementation by numerical simulation is the MCRG method. 
Alternatively, one may incorporate some judicious approximations in the definition of the 
transformation at the outset so that it becomes explicitly computable. 
%Such approximate block transformations can be highly effective for many purposes. 
Potential moving transformations are in this class. 

Partition a hypercubic $d$-dimensional lattice of spacing $a$ into $d$-dimensional hypercubes of linear size $ba$ ($b$ an integer). 
The potential moving procedure \cite{K} in the case of pure gauge theories consists of 
moving all plaquette interactions in the interior of each such hypercube to its $(d-1)$-dimensional 
boundary. After the move their strength is renormalized by some adjustable factor to 
compensate for the move. 
Next the plaquettes inside this $(d-1)$-dimensional boundary are moved to its $(d-2)$-dimensional boundary and similarly renormalized. Continuing this process to its conclusion one ends up with the system on a lattice of spacing $ba$ whose elementary plaquettes are tiled by at most $b^2$ plaquettes of the original lattice. 

This series of moves may be implemented in a number of somewhat different ways \cite{PM1}. 
The simplest choice though is to first perform all moves in sequence and with common plaquette  interaction renormalization $\zeta$ after each move. This is the scheme we adopt here. 
It results in isotropic couplings in all 
directions, and plaquettes on the blocked lattice (of spacing $ba$) tiled by exactly $b^2$ plaquettes of the original lattice; and with each of these tiling plaquettes renormalized by a total factor $\zeta^{(d-2)}$. The integrations over those bond variables belonging to the tiling plaquettes, and interior to the blocked lattice elementary plaquettes, can now be performed and renormalized.  
This completes the block step yielding the theory on the blocked lattice.

This procedure may be formulated as follows. Let, as usual, $U_b \in G$ denote the 
bond variables, and $U_p = \prod_{b\in \partial p} U_b$ their product around plaquette $p$. 
General elements of the gauge group $G$ are denoted by $U$. 
Let $A_p(U_p,n)$ denote a plaquette action on lattice of spacing $b^na$ 
and consider the character expansion  
\beq 
\exp \left(-A_p(U_p, n)\right) 
   = \sum_j\;d_j\, F_j(n)\,\chi_j(U_p) \;. \label{exp} 
\eeq
The sum is over all inequivalent irreducible representations labeled by $j$, with $\chi_j$ and $d_j$ denoting the character and dimension, respectively, of the $j$-th representation.  
%Thus, e.g., for $SU(2)$, $j=0,1/2,1,\ldots$ and $d_j=2j+1$. 
From (\ref{exp}), using orthogonality of characters, one has  
\beq
 F_j(n) = \int\, d U\;
\exp \left(-A_p(U,n)\right) \,{1\over d_j}\,\chi_j^*(U)\;,\label{Fourier}
\eeq
where $d U$ denotes Haar measure on $G$.  In this paper we consider only $G=SU(N), N=2,3$. 
The action itself 
is of course completely specified by the set of $F_j(n)$ coefficients and vice versa. It can be written in the general form 
\beq
A_p(U_p,n) = \sum_j \;{1\over  d_j} \beta_j(n) \, {1\over 2 l_j}[ \chi_j(U_p) + \chi_j(U_p^{-1})]     \label{actexp}
\eeq 
with $l_j=1$ for self-conjugate and $l_j=2$ for non-self-conjugate representations. (For   
$SU(2)$, in particular, $l_j=1$ for all $j$.)
In general we need consider actions of the form (\ref{actexp}) with any (infinite) number of characters, i.e. couplings $\beta_j$. It is useful to define an effective coupling $g^{(n)}$  characterizing a given action of the form (\ref{actexp}). With $\{t\}$ denoting the $SU(N)$ generators 
and $|\hat{m}|=1$, we let  
\beq
\beta^{(n)} = {2N\over g^{(n)\,2}} \equiv  \left. 2N {d^2 A_p(e^{i\theta \hat{m}\cdot t}, n) \over d \theta^2}\right|_{\theta=0} \, .\label{effcoupl}
\eeq 
(\ref{effcoupl}) is of course independent of the direction $\hat{m}$.  
In the perturbative regime this reduces to the usual definition of gauge coupling. In the non-perturbative regime any definition of  a `coupling' is of course some scheme-dependent choice. 
We adopt (\ref{effcoupl}) to track the RG evolution of (\ref{exp}), (\ref{actexp}); it 
provides, in particular, a good parametrization of the renormalized trajectory (see below).

A lattice block step $b^n a \to b^{n+1}a$  of the type described above can now be succinctly formulated as a prescription for obtaining the character expansion coefficients $F_j(n+1)$ in terms of the $ F_j(n)$: 
\beq 
F_j(n+1) = \left[ \int\, d U\;
\left[ \sum_i\;d_i\, F_i(n)\, \chi_i(U) \right]^{\zeta^{(d-2)}}
     \,{1\over d_j}\,\chi_j^*(U)  \right]^{r^2}  \;.   \label{recur} 
 \eeq
The inner bracket factor results from the symmetric potential moves described above. Subsequent 
boundary integrations modify each resulting 
expansion coefficient by a further amount controlled by the parameter $r$.  
Specification of $\zeta$ and $r$ completes the block step.  

Our decimation transformation $b^na \to b^{n+1}a$ is defined by (\ref{recur}) 
with \cite{PM2}: 
\bea
\zeta &  = & b\left[1 - c \, g^{(n)\,2} \right]          \label{decpar1} \\
r & = & b \left[ 1 - c \, g^{(n)\,2} \right]  \, ,      \label{decpar2}
\eea
where $c$ is an adjustable decimation parameter.  
For sufficiently large effective coupling values the $g^{(n)2}$ dependence 
in (\ref{decpar1}) - (\ref{decpar2}) has to be suitably leveled off \cite{sccutoff}, since we require 
$\zeta > 0$, $r>0$. This is, however, not explicitly indicated here as this regime is well outside the range of couplings encountered in our
applications of the recursions (\ref{recur}) below. 

In the following $c$, for given $b$, will be treated as 
a decimation parameter tuned for optimization of the procedure outlined in the next section. 
It is interesting to note, however, that this turns out to give values (section \ref{NumRes}) not far from the weak coupling computed values \cite{PM2, Kar}.    

It is convenient to work with normalized coefficients 
$f_j=F_j/F_0$ by factoring out the trivial representation coefficient in (\ref{exp}); the correspondingly normalized action differs by a trivial 
shift of the constant (trivial character) part in (\ref{actexp}). Effective couplings (\ref{effcoupl})
are also conveniently computed directly in terms of the $\{f_j\}$.

\section{Lattice matching of decimations \label{LM}} 
Given a $d$-dimensional lattice system with action $A(K)$  defined by a set of couplings 
$K = \{K_i\}$  RG block transformations by a scale factor $b$ generate a 
flow in action space: 
\beq 
K^{(0)} \to K^{(1)} \to K^{2)} \to \cdots \to K^{(n)} \to \cdots \,, \label{flow}
\eeq
where   $K^{(n)}=\{K^{(n)}_i\}$ denotes the couplings after $n$ blocking steps from the initial point $K^{(0)} \equiv K$.  
Since the physical correlation length remains of course fixed, the (dimensionless) lattice correlation length $\xi^{(n)}$ at step $n$ gets rescaled as  $\xi^{(n)} = 
\xi^{(0)}/b^n$.  
%Such an RG flow can have fixed points at $\hat{\xi}=0$ (strong coupling) or 
%$\hat{\xi}=\infty$ (critical). 

The resulting RG flow is toward a fixed point along irrelevant directions (couplings) and away from the fixed point along relevant directions (couplings). Irrespective of the starting point $K^{(0)}$
then, the flow, after a sufficient number of blocking steps, will approach the 
unique Wilsonian renormalized trajectory (RT) emanating from the fixed point along the relevant directions. 

Consider two sets of couplings $K$ and $K^\prime$. If the two RG flows starting from 
them end up at the same point on the RT after the same number of blocking steps $n$, then, since at the end point 
$\xi^{(n)}= \xi^{\prime\,(n)}$, the correlation lengths $\xi^{(0)}\equiv \xi$ and $\xi^{\prime\,(0)}
\equiv \xi^\prime$ 
at $K$ and $K^\prime$, respectively, must be equal; and since the physical correlation length is constant, $K$ and $K^\prime$ must also have the same lattice spacing $a$. 
By the same token, if the flows from $K$ and $K^\prime$ reach the same point on the RT after 
$n$ and $(n-m)$ steps, respectively, then the lattice correlation lengths at $K$ and $K^\prime$ must be related as  
\beq 
\xi^\prime = b^{-m} \xi \, ; \label{correl1}
\eeq
and the lattice spacings $a$ and $a^\prime$ as 
\beq 
a^\prime = b^m a \, . \label{latspace1}
\eeq

To identify such pairs of couplings we need to ascertain that, after $n$ and $(n-m)$ RG steps respectively,  the same point is reached on the RT. This can be done in two ways. One is 
to show that $A(K^{(n)}) = A(K^{\prime {(n-m)}})$.  This requires that one obtain the blocked action at each step. Another way is to show that the expectations of every operator, measured 
after performing the corresponding number of blocking steps from the initial two actions, agree. 
Either way, blocking $n$ times from a starting point $K$, and then adjusting another starting 
point $K^\prime$ so that after $(n-1)$, or, more generally, $(n-m)$ times, matching is achieved
is referred to as two-lattice matching \cite{H1}.   
%Identifying such pairs $K, K^\prime$ for each $n$, gives, in the large $n$ limit, the so-called bare step scaling function 
%\beq
%s_b(K,b) \equiv K - K^\prime \, . 
%\label{bstep1} 
%\eeq

If blockings  are performed numerically by MCRG, the second method appears  easier to use. Obtaining the blocked action can be difficult, whereas it is possible, at least in principle, to generate a
Boltzmann-weighted configuration ensemble for the blocked action by instead blocking the configurations of an ensemble generated from the original action. These can then be used to measure observables \cite{H2}. 
In practice, of course, due to lattice size limitations, only a rather small number of block steps is possible by MCRG, 
so getting close enough to the RT is not guaranteed. In this connection, since the location of the fixed point is block definition dependent, appropriate fine-tuning of free parameters in the block transformation definition can be crucial for achieving rapid approach in few steps.

In this paper we employ two-lattice matching with RG block transformations implemented by the recursions (\ref{recur}) described above. They can be explicitly evaluated to any desired accuracy on lattices of any size, so no inherent limitations due to finite size arise. 
The blocked action resulting after each RG step is explicitly obtained, so it can used to 
ascertain approach to the RT and perform two-lattice matching. The transformations contain one parameter (cf. (\ref{decpar1}) - (\ref{decpar2})), which, as already mentioned, 
is fixed for optimization of the matching.

In the following the starting action ($n=0$) will always be taken to be the fundamental representation Wilson action. 
%In the notation of equation (\ref{actexp}) then this corresponds to 
%$\beta_{\rm f}(0)=\beta_{\bar{\rm f}}(0)=\beta$, and $\beta_j(0)=0$ for all other $j$. 
Other choices such as mixed actions containing 
several representations can be treated in exactly the same way. The flow under successive decimations reaches a unique RT irrespective of such a choice, though of course the number of steps needed to reach it depends on the initial point in action space. 
The important feature characterizing these decimations is that, regardless of the choice of the 
initial plaquette action, a single step suffices to generate an action 
of the form (\ref{actexp}) generally containing the full (infinite) set of representations. 
Flow in such an infinite-dimensional interaction space makes it possible to avoid getting stuck at 
(finite-dimensional) lattice artifact boundaries.  

\begin{figure}[ht]
\begin{center}
\includegraphics[width=0.8\textwidth]{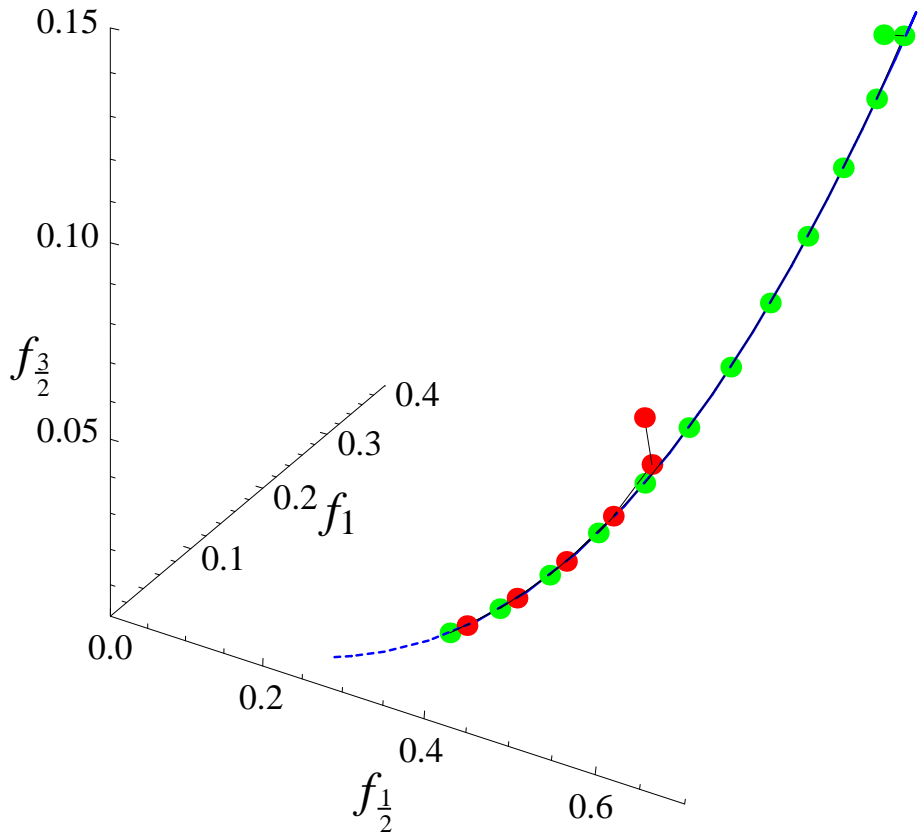}
\end{center}
\caption{RG flow and rapid approach to a unique renormalized trajectory starting from the 
$SU(2)$ fundamental representation Wilson action with $\beta=4$ (green dots) and $\beta=2.5$ (red dots). The first three non-trivial (normalized) expansion coefficients in (\ref{exp}) are shown.   \label{RGf1}}
\end{figure}

With the fundamental representation Wilson action as the starting 
point we  find that the approach to the unique RT is very rapid; it generally takes only two steps to get to it. This is illustrated in Figure \ref{RGf1}. 

A good way to parametrize points along the RT is by the effective coupling (\ref{effcoupl}) of the action corresponding to each RT point.  If, starting from some 
Wilson action coupling $\beta$, after $n$ steps the point $\beta^{(n)}(\beta)$ lies on the RT, subsequent RG steps generate a sequence of points $\beta^{(n+1)}$, $\beta^{(n+2)}, \cdots$ 
hopping along the RT.  With scale factor $b=2$, and for all large and intermediate values of $\beta^{(n)}$, the effective beta function is varying slowly enough 
for a linear interpolation to provide an excellent approximation to the RT points lying between pairs of neighboring points $\beta^{(n)}, \beta^{(n+1)}$.  
So we write 
\beq
\beta^{(n+h)} = \beta^{(n)} + h (\beta^{(n+1)} - \beta^{(n)}) \,, \qquad   0 <  h < 1 
\, .\label{interpol}
\eeq 
This may be viewed as performing $n$ steps followed by a step with fractional scale 
factor to reach a point $\beta^{(n+h)}$ lying between point $\beta^{(n)}$ and 
$\beta^{(n+1)}$ on the RT. (The block transformation rules (\ref{recur})-(\ref{decpar2}) can 
indeed be formally extended to non-integer scale factor, but we need not make actual use of this 
here.)  
(\ref{interpol}) asserts that the location of this intermediate point is accurately given by 
linear interpolation.

Though explicitly computable to any accuracy, our decimations are of course 
approximate. They become exact in the strict $\beta\to \infty$ limit. Computation of the step scaling function (beta function) from the decimations in the weak coupling scaling region reproduces the 
perturbation theory prediction to within $2\%-3\%$.  
The next question to be probed by further computation then is how close an approximation these 
decimations give to the exact RT in the transition to the non-perturbative regime.  
MCRG construction of blocked actions \cite{T1mcrg} shows that one-plaquette terms 
with a large number of characters are the most relevant action terms for long-scale dynamics. 
This is precisely the type of action resulting from our decimations.

In the following two-lattice matching of our RG decimations is used to obtain 
critical couplings and string tensions for the $SU(2)$ and $SU(3)$ gauge theories.

\subsection{Critical couplings \label{cc}}  Consider the $(3+1)$-dimensional lattice theory at physical temperature $T$. 
Since $T=1/aN_\tau$ for lattice of time extent $N_\tau$ and spacing $a$, lattice of extent 
$N^\prime_\tau$, spacing $a^\prime$ is related by  
\beq
a^\prime = {N_\tau \over N^\prime_\tau } a \,. \label{latspace2}
\eeq
If after blocking the two lattices $n$ and $n^\prime$ times, respectively, the two flows 
reach the same point on the RT trajectory, (\ref{latspace1}) and (\ref{latspace2}) imply 
\beq
n - n^\prime = \log_b \left({N_\tau \over N^\prime_\tau}\right) \,. \label{latspace3} 
\eeq
If, in particular, $T=T_c$, one has 
\beq 
\beta^{(n)}(\beta_c(N_\tau)) = \beta^{(n^\prime)} (\beta_c(N^\prime_\tau)) \,.\label{eqeffcr}
\eeq 

(\ref{latspace3}) and (\ref{eqeffcr}) afford a straightforward way of obtaining critical 
couplings by matching once one such coupling is known. 
Assuming $\beta_c(N_\tau)$ known, it is convenient to simply choose 
\bea
n& = & \log_b N_\tau   \nonumber \\
n^\prime & = & \log_b N^\prime_\tau \label{nchoice}
\eea 
so that (\ref{latspace3}) is satisfied. $n$, $n^\prime$ must be large enough to be on the RT, but this is not a problem since one can always simply add a common  integer to the r.h.s. of both equations in (\ref{nchoice}). Also, note that the so-chosen $n$ or/and $n^\prime$ may turn out to be non-integer. In such a case, one performs $[n]$ and $[n]+1$ steps, where $[n]$ is the nearest integer to the chosen $n$ from below, and uses (\ref{interpol}) - and similarly for $n^\prime$.  
With $n$, $n^\prime$ and $\beta_c(N_\tau)$ given, (\ref{eqeffcr}) is then solved for $\beta_c(N^\prime_\tau)$, i.e., the starting point of the flow on the $N^\prime_\tau$ lattice is adjusted 
to satisfy (\ref{eqeffcr}).  

\subsection{String tensions  \label{st}} The string tension $\hat{\sigma}(\beta) = a ^2\sigma$ is another quantity that can be 
obtained at different couplings by the same method. Consider two RG flows with starting point 
the Wilson action at couplings $\beta_0$ and $\beta_1$ and ending up at the same point on the RT after $n_0$ and $n_1$ steps, respectively. Then 
\beq 
\beta^{(n_0)}(\beta_0) = \beta^{(n_1)}(\beta_1)  \label{eqeffst} 
\eeq 
and 
\beq
a_1\sqrt{\sigma} = b^{(n_0 - n_1)} a_0 \sqrt{\sigma}  \label{eqst}
\eeq 
by (\ref{latspace1}). 

Suppose we know $\hat{\sigma}(\beta_0)$. Choose $n_0$ large enough to be on the RT. 
Then $n_1$ is fixed so that (\ref{eqeffst}) is satisfied. In practice this is done by 
performing $n$ and $n+1$ decimation RG steps 
with initial coupling $\beta_1$, where $n$ is determined so that  
\beq 
\Big(\beta^{(n_0)}(\beta_0) - \beta^{(n+1)}(\beta_1) \Big)\Big(\beta^{(n_0)}(\beta_0) - \beta^{(n)}(\beta_1) \Big) 
\leq 0 \;. \label{bsteps} \nonumber 
\eeq
Then, by (\ref{interpol}), $n_1$  satisfying (\ref{eqeffst}) is given by $n_1=n+h$ with 
\beq 
h= { \beta^{(n_0)}(\beta_0) - \beta^{(n)}(\beta_1)\over \beta^{(n+1)}(\beta_1) - \beta^{(n)}(
\beta_1) }     \, . \label{eqeffinterpol} 
\eeq
$\sqrt{\hat{\sigma}}(\beta_1) = a_1\sqrt{\sigma}$ is then obtained directly from (\ref{eqst}).

\section{Results for critical couplings and string tensions \label{NumRes}} 

 For $SU(2)$ we typically use fifty group characters in the expansions (\ref{exp}). This implies 
 for, say, $\beta=5$ omitted higher character coefficients $f_j=F_j/F_0$, and accompanying bounds on the series remainder, of the order of $10^{-45}$. For $SU(3)$ we truncate (\ref{exp}) at characters $j\equiv (p,q)$ with $p\geq 20$, $q\geq 20$; this implies remainders at $\beta=10$ of less than $10^{-12}$. Iteration under (\ref{recur}) 
results into decreasing expansion coefficients. Errors due to truncation in the character expansions (\ref{exp}) are thus totally negligible.  

The scale factor is always taken to be $b=2$. The only adjustable parameter in the decimation recursions (\ref{recur}) - (\ref{decpar2}) is $c$ which is tuned for optimized matching. We set $c=0.10$ in the case of $SU(2)$ and $c=0.24$ in the case of $SU(3)$. 
With no other parameters present, straightforward numerical evaluation of the recursion relations can then be implemented.

\begin{table}[htb]
\centering
\begin{tabular}{|c|@{\hspace{0.5cm}}c@{\hspace{0.5cm}}|@{\hspace{0.5cm}}c@{\hspace{0.5cm}}|@{\hspace{0.5cm}}c@{\hspace{0.5cm}}|}
  \hline
  % after \\: \hline or \cline{col1-col2} \cline{col3-col4} ...
  $N_\tau$ &  $\beta_c$ & $\beta_c$&$\beta_c$(MC) \\
  \hline
  3  & 2.1875 & 2.1957& 2.1768(30) \\
  4  & 2.2909 & \underline{2.2991}& 2.2991(02) \\
  5 & 2.3600 & 2.3683 & 2.3726(45)  \\
  6 & 2.4175 & 2.4258  & 2.4265(30) \\
  8 & 2.5097 & 2.5180 & 2.5104(02) \\
  12  & \underline{2.6355} & 2.6440 & 2.6355(10) \\
  16  & 2.7275 & 2.7361 & 2.7310(20)\\
  32  & 2.9487 & 2.9574   &   \\
  \hline
\end{tabular}
\caption{Critical couplings $\beta_c(N_\tau)$ for $SU(2)$ computed from lattice 
matching of decimations. Column 1 and 2 show the values obtained for two different choices 
(underlined entries) of the one data point taken from MC data (see text). Column 3 shows the values 
from MC simulations for comparison. \label{TablebetacrSU(2)} } 
\end{table}
%\begin{figure}
%  \centering{}\includegraphics[width=5in]{CriticalCouplingSU(2).png}
%  \linespread{1.5}\caption{$SU(2)$ critical couplings from table.\ref{TablebetacrSU(2)} \label{betacrSU(2)fig}}
%\end{figure}

\begin{table}[htb]
\centering
\begin{tabular}{|c|@{\hspace{0.5cm}}c@{\hspace{0.5cm}}|@{\hspace{0.5cm}}c@{\hspace{0.5cm}}|@{\hspace{0.5cm}}c@{\hspace{0.5cm}}|}
%{|c|l|l|l|}
 \hline
  % after \\: \hline or \cline{col1-col2} \cline{col3-col4} ...
 $N_\tau$ & $\beta_c$ & $\beta_c$ & $\beta_c$(MC) \\
 \hline
  4  & 5.6501 & 5.6329 & 5.6925(002) \\
  6 & \underline{5.8941} & 5.8773  & 5.8941(005) \\
  8 & 6.0773 & 6.0595 & 6.0010(250),6.0625(18)  \\
  10  & 6.2018 & 6.1837 & 6.1600(70)\\
  12 & 6.3084 & 6.2900  & 6.2680(120),6.3385(55) \\
  14  & 6.4015 & \underline{6.3830} & 6.3830(100)\\
  16  & 6.4845 & 6.4658 & 6.4500(500)\\
  32  & 6.9024 & 6.8829 &  \\
 \hline
\end{tabular}
\caption{Critical couplings $\beta_c(N_\tau)$ for $SU(3)$ 
computed from lattice matching of decimations and comparison with MC simulation data. 
Same format as in Table \ref{TablebetacrSU(2)}. }
\label{TablebetacrSU(3)}
\end{table}

We take one value of $\beta_c(N_\tau)$ from MC data, which serves to fix the 
scale. We then apply the procedure of section \ref{cc} to obtain critical coupling values 
for other lattices. Results for $SU(2)$ are shown in Table \ref{TablebetacrSU(2)}. Two sets of computed $\beta_c$ values are shown  in 
Table \ref{TablebetacrSU(2)} (columns 1 and 2) corresponding to two different choices of the MC data point (underlined entries). The table also shows comparison with the values obtained by MC simulation \cite{LTW} - \cite{Hetal}, \cite{Phetal}  in each case (column 3). The agreement is remarkably good - typically 
of the order of $1\%-2\%$. 

Results for critical couplings in the $SU(3)$ gauge theory are displayed in Table \ref{TablebetacrSU(3)}.  
Agreement with MC simulation data  \cite{LTW}, \cite{Phetal} 
% \cite{Langelage2.1,Lucini3.1}. 
is again very good, typically within a few 
percent.

String tensions in $SU(2)$ obtained by the method of section \ref{st} are displayed in the same format in Table \ref{TableSTSU(2)}. Again, two sets of values are shown (columns 1 and 2)  
corresponding to two different choices (underlined entries) of the MC data point used as initial input.  
The corresponding results in the case of $SU(3)$ are shown in Table \ref{TableSTSU(3)}. 
Good agreement with MC data \cite{Hetal} - \cite{EHK} 
is again obtained in all cases. 

\begin{table}[h]
\centering
\begin{tabular}{|c|@{\hspace{0.5cm}}c@{\hspace{0.5cm}}|@{\hspace{0.5cm}}c@{\hspace{0.5cm}}|@{\hspace{0.5cm}}c@{\hspace{0.5cm}}|}
%{|l|l|l|l|}
  \hline
  % after \\: \hline or \cline{col1-col2} \cline{col3-col4} ...
  $\beta$  & $a\sqrt \sigma $& $a\sqrt \sigma $ & $a\sqrt \sigma $(MC) \\
  \hline
  2.2  & 0.5019 & 0.5161 & 0.4690(100)\\
  2.3  & 0.3654 & 0.3756   & 0.3690(30)\\
  2.4 & 0.2619 & 0.2696 & 0.2660(20) \\
  2.5  & 0.1903 & 0.1957  & 0.1905(08)\\
  2.5115& \underline{0.1836} & 0.1888  & 0.1836(13) \\
  2.6 & 0.1373 & 0.1415   & 0.1360(40) \\
  2.7 & 0.1002 & 0.1031 & 0.1015(10) \\
  2.74 & 0.0884 & \underline{0.0911} & 0.0911(08) \\
  2.85  & 0.0622 & 0.0641 & 0.0630(30)\\
  \hline
\end{tabular}
\caption{String tensions $a\sqrt \sigma $ for $SU(2)$ computed from lattice 
matching of decimations. Column 1 and 2 show the values obtained for two different choices 
(underlined entries) of the one data point taken from MC data (see text). Column 3 shows the values 
from MC simulations for comparison.}\label{TableSTSU(2)}
\end{table}

\begin{table}[ht]
\centering
\begin{tabular}{|c|@{\hspace{0.5cm}}c@{\hspace{0.5cm}}|@{\hspace{0.5cm}}c@{\hspace{0.5cm}}|@{\hspace{0.5cm}}c@{\hspace{0.5cm}}|}
%{|l|l|l|l|}
  \hline
  % after \\: \hline or \cline{col1-col2} \cline{col3-col4} ...
  $\beta$ & $a\sqrt \sigma $ & $a\sqrt \sigma $  & $a\sqrt \sigma $(MC) \\
  \hline
  5.54 & 0.5580 & 0.5878  & 0.5727(52) \\
  5.6  & 0.5070 & \underline{0.5295}  & 0.5295(09), 0.5064(28)\\
  5.7  & 0.4205 & 0.4264  & 0.4099(12), 0.3879(39)\\
  5.8 & 0.3486 & 0.3508   & 0.3302(15) \\
  5.9  & 0.2919 & 0.2931 & 0.2702(19) \\
  6.0   & 0.2465 & 0.2433  & 0.2269(62), 0.2209(23)\\
  6.2  & 0.1698 & 0.1671  & 0.1619(19), 0.1604(11)\\
  6.4 & \underline{0.1214} & 0.1180  & 0.1214(12), 0.1218(28) \\
  6.5 & 0.1010 & 0.0983  & 0.1068(09) \\
  6.8 & 0.0616 & 0.0599  & 0.0738(20) \\
  \hline
\end{tabular}
\caption{String tensions $a\sqrt \sigma $ for $SU(3)$ computed from lattice matching of decimations. Same format as in Table \ref{TableSTSU(2)}.} \label{TableSTSU(3)}
\end{table}

\section{Conclusions \label{Concls}} 
The RG decimation recursion relations given in section \ref{LM} were used in conjunction with 
two-lattice matching to compute critical couplings and string tensions in $SU(2)$ and $SU(3)$ pure lattice gauge theories. The decimations contain only one adjustable parameter that was fixed, in the case of each group, to an optimized value given in section \ref{NumRes}. Using one initial value obtained by MC simulation, critical couplings and string tensions  were then obtained for a variety of other lattices by lattice matching of our decimations. 
The results were found to be in very good agreement with those obtained by MC simulation. 
The method evidently provides a cheap way of quickly obtaining accurate predictions for these quantities for a wide range of lattice sizes. 

Critical couplings and string tensions are quantities pertaining to long-distance non-perturbative dynamics. The actions evolving under the decimations are plaquette actions with a large 
(infinite) number of representations. As mentioned above, MCRG constructions of blocked actions \cite{T1mcrg} indicate that these are the action terms most relevant for long distance dynamics. This may be one reason underlying the method's apparent efficacy.  

There are two directions in which this work could be further pursued. One is to consider more 
general block transformations. The decimations employed here may indeed be viewed as special cases of more elaborate blocking schemes. These will, in general, involve additional decimation parameters, but are likely necessary for computation of observables over different length scales. 
The other direction is the inclusion of fermions. Block transformations involving fermions 
present a generally much harder problem. Use of relatively simple block schemes in a somewhat different context, however, have given some early encouraging results \cite{CT}. 

\vspace{0.5cm}
This work was partially supported by the NSF under NSF-PHY-0852438.


\begin{thebibliography}{99}
\bibitem{PM1}  Every time a move is performed a certain subset of integrations over the bond variables becomes exactly doable.  
One may elect to perform or not perform these integrations before the next move. Performing 
integrations before moves generally results into anisotropic couplings for the resulting blocked 
lattice action. Also, different renormalization parameters may be chosen for each 
move. Even more elaborate schemes can be devised introducing more decimation parameters.
\bibitem{PM2} The form (\ref{decpar1}) - (\ref{decpar2}) may be motivated as follows. 
Whenever a plaquette move is performed in one given direction, a lattice spacing anisotropy by a factor $b$ is created relative to the other directions. At weak coupling the plaquette strength gets renormalized by a factor $\zeta$ of precisely the form (\ref{decpar1}) where $c=c(b)$ can be obtained analytically to 1-loop \cite{Kar}. We apply this correction with common $c$ in all directions so as to arrive at the isotropic decimation rule (\ref{recur}). More involved rules involving different 
$c(b)$ in different directions, and indeed different choices for $\zeta$ and $r$, are possible 
(cf. \cite{PM1});  such more general decimation schemes, however, will not be considered here. 
Going in the other direction, we note that the choices $\zeta=b$ and $r=b$ define the original MK decimations \cite{K, M} which correspond to simply increasing plaquette interaction strengths by a factor equal to the number of plaquettes displaced by each move   
without taking any other quantum corrections into account.
\bibitem{sccutoff} Once within the strong coupling regime, dependence on  the precise form of such a cutoff is in fact rather weak.
\end{thebibliography}
\end{document}